\documentclass[epj]{svjour}
\usepackage{graphicx}
\usepackage{multirow}
\usepackage{graphics}
\usepackage{hyperref}
\hypersetup{colorlinks = true, allcolors = blue}
\usepackage{authblk}
\usepackage{hepparticles}
\usepackage{hepunits}
\usepackage{hepnames}

\providecommand{\MGaMC}{{\sc\small MadGraph5\_aMC@NLO}}

\begin{document}
\title{A special Higgs challenge}
\subtitle{Measuring the mass and production cross section with ultimate precision at FCC-ee}
\author{%Patrizia Azzi\inst{1} \and 
Paolo Azzurri\inst{1} \and Gregorio Bernardi\inst{2} \and Sylvie Braibant\inst{3} \and David d'Enterria\inst{4} \and Jan Eysermans\inst{5} \and Patrick Janot\inst{4}  \and Ang Li\inst{2} \and Emmanuel Perez\inst{4}% etc
% \thanks is optional - remove next line if not needed
%\thanks{\emph{Present address:} Insert the address here if needed}%
}                     % Do not remove
\offprints{}          % Insert a name or remove this line
\institute{%INFN, Sezione di Padova, Italy \and 
INFN, Sezione di Pisa, Italy \and CNRS/IN2P3, APC Paris, France \and Bologna University and INFN, Sezione di Bologna, Italy \and CERN, EP Department, Geneva, Switzerland \and MIT, Cambridge, Massachusetts, USA} 
\date{%Received: \today / Revised version: \today 
{\sl 
(Submitted to EPJ+ special issue:
A future Higgs and Electroweak factory (FCC): Challenges towards discovery, Focus on FCC-ee)}}
% The correct dates will be entered by Springer
%
\abstract{
%Insert your abstract here.
The FCC-ee offers powerful opportunities to determine the Higgs boson parameters, exploiting over $10^6$ ${\rm e^+e^- \to ZH}$ events and almost $10^5$ ${\rm WW \to H}$ events at centre-of-mass energies around 240 and 365\,GeV. This essay spotlights the important measurements of the ZH production cross section and of the Higgs boson mass. The measurement of the total ZH cross section is an essential input to the absolute determination of the HZZ coupling -- a ``standard candle'' that can be used by all other measurements, including those made at hadron colliders -- at the per-mil level. A combination of the measured cross sections at the two different centre-of-mass energies further provides the first evidence for the trilinear Higgs self-coupling, and possibly its first observation if the cross-section measurement can be made accurate enough. The determination of the Higgs boson mass with a precision significantly better than the Higgs boson width (4.1\,MeV in the Standard Model) is a prerequisite to either constrain or measure the electron Yukawa coupling via direct ${\rm e^+e^- \to H}$ production at $\sqrt{s} = 125$\,GeV. %Traditionally, both the cross section and the mass are measured with the H$\ell^+\ell^-$ final state independently of the Higgs boson decay. These measurements are statistically limited, to 0.5\% for the cross section and 5\,MeV for the mass, by the small Z leptonic branching fraction. 
Approaching the statistical limit of 0.1\% and $\mathcal{O}(1)$\,MeV on the ZH cross section and the Higgs boson mass, respectively, sets highly demanding requirements on accelerator operation (ZH threshold scan, centre-of-mass energy measurement), detector design (lepton momentum resolution, hadronic final state reconstruction performance), theoretical calculations, %(higher-order corrections to the cross section), 
and analysis techniques (efficiency and purity optimization with modern tools, constrained kinematic fits, control of systematic uncertainties). These challenges are examined in turn in this essay.
 % end of PACS codes
} %end of abstract
\authorrunning{P. Azzurri et al.}
\titlerunning{Measuring the Higgs mass and production cross section with ultimate precision at FCC-ee}
\maketitle

\section{Introduction: State of the art}
\label{section:intro}

\begin{figure}[htbp]
\centering
\begin{tabular}{ccc}
\begin{minipage}[b]{0.15\textwidth}
\centering
{\small Higgsstrahlung}

\vspace{2.4cm}
{\small WW fusion}

\vspace{1.2cm}

\vspace{0.5cm}
\end{minipage}
\begin{minipage}[b]{0.18\textwidth}
\centering
\includegraphics[width=\textwidth]{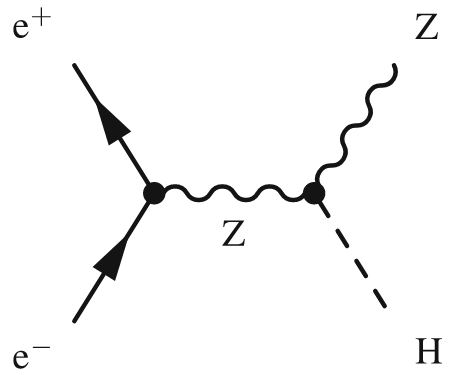}
\includegraphics[width=\textwidth]{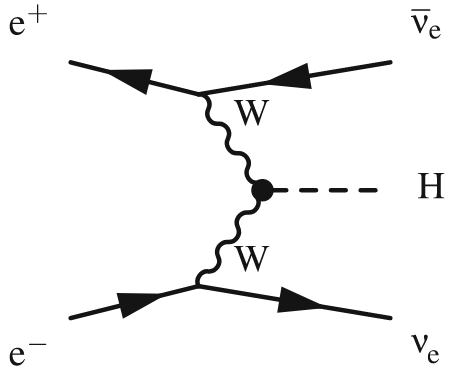}

\vspace{0.5cm}
\end{minipage}
\begin{minipage}[b]{0.60\textwidth}
\includegraphics[width=\textwidth]{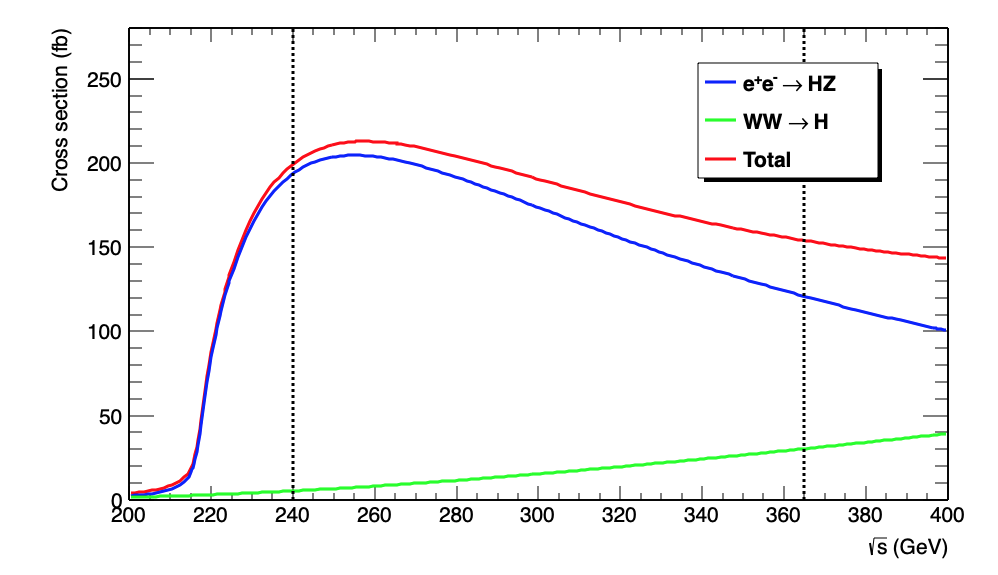}
\end{minipage}
\end{tabular}
\caption{\small (Left) Feynman diagrams for the Higgsstrahlung (top) and the WW fusion (bottom) processes. (Right) Improved-Born Higgs production cross sections (with initial state radiation included~\cite{Berends:1984dw}), as predicted by {\tt HZHA}~\cite{Altarelli:300671} as a function of the centre-of-mass energy for $m_{\rm H} = 125$\,GeV. The small interference term between the two diagrams in the $\rm H \nu_e \bar\nu_e$ final state is included in the WW fusion cross section. Vertical dashed lines indicate the $\sqrt{s}$ values foreseen at FCC-ee.}
\label{fig:HiggsProduction}
\end{figure}

In $\rm e^+e^-$ collisions at centre-of-mass energies ($\sqrt{s}$) from 200 to 400\,GeV, the two main Higgs production mechanisms are the Higgsstrahlung process, $\rm e^+e^- \to ZH$, and the $\rm WW$ fusion process, $\rm e^+e^- \to H\nu_e\bar\nu_e$, with Feynman diagrams and cross sections shown in Fig.~\ref{fig:HiggsProduction}. In the baseline run plan~\cite{Blondel:2021ema} and with two interaction points (IPs), over one million ZH events and almost 100,000 $\rm WW \to H$ events Higgs bosons will be produced at FCC-ee with $\sqrt{s}$ between 240 and 365\,GeV. The design of a layout with four IPs is under study, and would increase these figures by a  factor 1.7 in the same running time.
The total ZH cross section $\sigma_{\rm ZH}$ can thus be determined in principle with an ultimate statistical precision of 0.1\%, should the ZH event selection be 100\% efficient and pure. In addition, the selection can be made independent of the Higgs boson detailed properties by counting events with an identified Z boson, and for which the mass recoiling against the Z clusters around the Higgs boson mass (Section~\ref{section:x-section}). %proportional to the square of the Higgs coupling to the Z boson ($g_{\rm HZZ}$), can therefore be determined at FCC-ee in a model-independent manner with a statistical precision of 0.1\%, by counting all events with an identified Z and for which the mass $m_{\rm recoil}$ recoiling against the Z, given by $m_{\rm recoil}^2 = s +m_{\rm Z}^2 - 2\sqrt{s}E_{\rm Z}$, clusters around the Higgs boson mass $m_{\rm H}$. 
An absolute measurement of $g_{\rm HZZ}$, unique to $\rm e^+e^-$ colliders, is therefore at hand with a statistical precision of half a per mil at FCC-ee. The position of the recoil mass peak also provides in turn an accurate measurement of the Higgs boson mass. Once $g_{\rm HZZ}$ has been determined, the measurement of the cross sections for each exclusive Higgs boson decay, $\rm H \to X{\overline X}$, 
\begin{equation}
    \sigma_{\rm ZH} \times \mathcal{B}({\rm H \to X\overline X}) \propto \frac{g^2_{\rm HZZ} \times g^2_{\rm HXX}}{\Gamma_{\rm H}}
    {\rm \ \ and\ \ } \sigma_{\rm H \nu_e\bar\nu_e} \times \mathcal{B}({\rm H \to X\overline X}) \propto \frac{g^2_{\rm HWW} \times g^2_{\rm HXX}}{\Gamma_{\rm H}},
\label{eq:crossb}
\end{equation}
gives access to all other couplings in a model-independent, absolute, way. For example, the ratio of the WW-fusion-to-Higgstrahlung cross sections for the same Higgs boson decay, proportional to $g^2_{\rm HWW}/g^2_{\rm HZZ}$,  yields $g_{\rm HWW}$, and the Higgsstrahlung rate with the $\rm H \to ZZ$ decay, proportional to $g^4_{\rm HZZ}/\Gamma_{\rm H}$, provides a determination of the Higgs boson total decay width $\Gamma_{\rm H}$. The measurement of $g_{\rm HZZ}$, and thus of the total ZH cross section, is a cornerstone of the Higgs physics programme at FCC-ee. Conservative values for the statistical precision on inclusive and exclusive ZH cross sections, obtained from preliminary FCC-ee conceptual studies with realistic beam and detector parameters~\cite{Benedikt:2651299}, are indicated in Table~\ref{tab:HiggsMeasurements}, and the resulting accuracy of Higgs couplings obtained from global fits to the FCC-ee measurements (the details of which are explained in Ref.~\cite{deBlas:2019rxi}), are listed in Table~\ref{tab:kappaEFT}. 

\begin{table}[h!]
\parbox{.45\linewidth}{
\centering
\caption{From Ref.~\cite{Benedikt:2651299}: Relative uncertainty (in \%) on $\sigma_{\rm ZH} \times \mathcal{B} ({\rm H \to X \overline X})$ and $\sigma_{\rm \nu_e\bar\nu_e H} \times \mathcal{B} ({\rm H \to X \overline X})$,  as expected from the FCC-ee data at 240 and 365\,GeV. \label{tab:HiggsMeasurements}}
%\vspace{2mm}
\begin{tabular}{l|r r|r r}
\hline %\hline
$\sqrt{s}$ & \multicolumn{2}{c|}{$240$\,GeV} & \multicolumn{2}{c}{$365$\,GeV} \\ \hline
Integrated luminosity & \multicolumn{2}{c|}{5\,${\rm ab}^{-1}$} & \multicolumn{2}{c}{$1.5$\,${\rm ab}^{-1}$} \\ \hline
Channel & ZH & $\rm \nu_e\bar\nu_e$ H &  ZH & $\rm \nu_e\bar\nu_e$ H \\  \hline
${\rm H \to any}$          & $\pm 0.5$ &           & $\pm 0.9$    &             \\ %\hline
${\rm H \to b \bar b}$     & $\pm 0.3$ & $\pm 3.1$ & $\pm 0.5$    & $\pm 0.9$   \\ %\hline
${\rm H \to c \bar c}$     & $\pm 2.2$ &           & $\pm 6.5$    & $\pm 10$    \\ %\hline
${\rm H \to gg}$           & $\pm 1.9$ &           & $\pm 3.5$    & $\pm 4.5$   \\ %\hline
${\rm H \to W^+W^-}$       & $\pm 1.2$ &           & $\pm 2.6$    & $\pm 3.0$   \\ %\hline
${\rm H \to ZZ}$           & $\pm 4.4$ &           & $\pm 12$     & $\pm 10$    \\ %\hline
${\rm H \to}$ $\tau^+\tau^-$    & $\pm 0.9$ &           & $\pm 1.8$    & $\pm 8$     \\ %\hline
$\rm H \to \gamma \gamma$ & $\pm 9.0$ &           & $\pm 18$     & $\pm 22$    \\ %\hline
${\rm H \to}$ $\mu^+\mu^-$   & $\pm 19$  &           & $\pm 40$     &             \\ %\hline
${\rm H \to invisible}$       & $<0.3$    &           & $<0.6$       &             \\ \hline %\hline
\end{tabular} 
}
\hfill
\parbox{.45\linewidth}{
\centering
\caption{\small From Ref.~\protect\cite{deBlas:2019rxi}: Precision on a few Higgs couplings $g_{\rm HXX}$ and on the total width $\Gamma_{\rm H}$ at FCC-ee, in the $\kappa$ framework and in a global Effective Field Theory fit.  
%The last row lists the precision expected on the total width $\Gamma_{\rm H}$.
%\vspace{3mm}
}
\label{tab:kappaEFT}
%\vspace{2mm}
%\resizebox{\columnwidth}{!}{
\begin{tabular}{l|c}
%\multicolumn{6}{c}{} \\
\hline Coupling \;\;\;\;& Precision (\%)  \\ %\hline
                        & ($\kappa$ framework / EFT) \\ \hline
$g_{\rm HZZ}$ &  0.17 / 0.26  \\ %\hline
$g_{\rm HWW}$ &  0.41 / 0.27   \\ %\hline
$g_{\rm Hbb}$ &  0.64 / 0.56 \\ %\hline
$g_{\rm Hcc}$ &  1.3 / 1.2 \\ %\hline
$g_{\rm Hgg}$ &  0.89 / 0.82  \\ %\hline
$g_{\rm H\tau\tau}$ & 0.66 / 0.57 \\ %\hline
$g_{\rm H\mu\mu}$ & 3.9 / 3.8 \\ %\hline
$g_{\rm H\gamma\gamma}$  & 1.3 / 1.2 \\ %\hline
$g_{\rm HZ\gamma}$ & 10. / 9.3 \\ %\hline
$g_{\rm Htt}$ & 3.1 / 3.1 \\ \hline
$\Gamma_{\rm H}$ & 1.1  \\ \hline
%$\mathcal{B}_{\rm inv}$ & $< 0.19$ \\ \hline
%$\mathcal{B}_{\rm EXO}$ & $< 1.0$ \\ \hline
\end{tabular}
%}
}
\end{table}

\begin{figure}[htbp]
\begin{center}
%\quad
\raisebox{0.1\height}{\includegraphics*[width=0.50\textwidth]{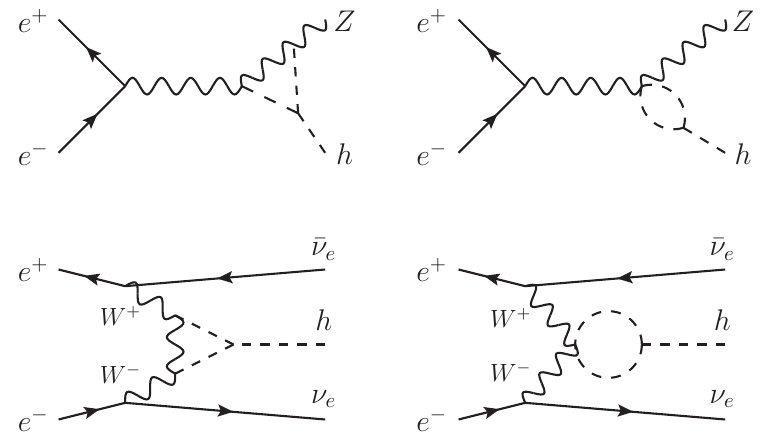}}
\includegraphics*[width=0.42\textwidth]{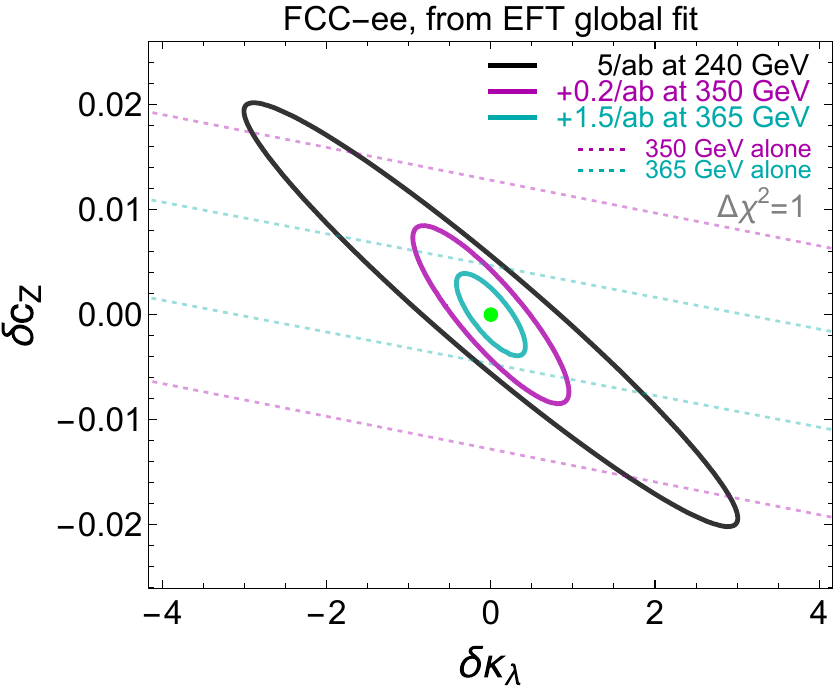}
\end{center}
\caption{From Ref.~\protect\cite{DiVita:2017vrr}. Left: Sample next-to-leading-order Feynman diagrams for single Higgs production involving the Higgs self-coupling. Right: Relative precision in the simultaneous determination of the Higgs self-coupling (here denoted $\kappa_\lambda$) and the HZZ/HWW coupling (here denoted $c_{\rm Z}$) at FCC-ee, with 240\,GeV (black ellipse), 350\,GeV (purpled dashed), and 365\,GeV (green dashed) data, and by combining data at 240 and 350 GeV (purple ellipse), and at 240, 350, and 365\,GeV (green ellipse).}
\label{fig:h3-TGC-atEE}
\end{figure}

The precise measurement of the ZH cross section can also give access to the Higgs boson self-coupling $g_{\rm HHH}$ via loop diagrams (shown in the left panel of Fig.~\ref{fig:h3-TGC-atEE}) as was realised for the first time in Ref.~\cite{McCullough:2013rea}. Indeed, the contribution of these diagrams to the ZH cross section amounts to $\sim$2\% at 240\,GeV and $\sim$0.5\% at 365\,GeV~\cite{DiMicco:2019ngk}, similar to or significantly larger than the experimental precision expected at FCC-ee. The dependence of the ZH cross section on the centre-of-mass energy allows in addition the $g_{\rm HZZ}$ and $g_{\rm HHH}$ couplings to be determined separately in a robust and model-independent manner~\cite{DiMicco:2019ngk,DiVita:2017vrr}, with a relative precision illustrated in the right panel of Fig.~\ref{fig:h3-TGC-atEE}. The corresponding contribution to the WW fusion cross section is almost independent of $\sqrt{s}$~\cite{DiVita:2017vrr,Maltoni:2018ttu}, which offers a complementary constraint on $g_{\rm HHH}$. No meaningful constraint can be obtained with one single centre-of-mass energy. Altogether, with the $\sigma_{\rm ZH}$ accuracy given in Table~\ref{tab:HiggsMeasurements}, a precision of $\pm 33\%$ can be achieved at FCC-ee on the Higgs self-coupling, reduced to $\pm 24\%$ with four interaction points (and therefore, four detectors) instead of two~\cite{DiMicco:2019ngk,Blondel:2018aan}. The first 3-to-4\,$\sigma$ evidence of the existence of the Higgs self-coupling is therefore within reach in 15 years at FCC-ee, a unique feature among all low-energy $\rm e^+e^-$ EW and Higgs factories currently contemplated.

It is important to note here that the $\sigma_{\rm ZH}$ accuracy of 0.5\% given in Table~\ref{tab:HiggsMeasurements} for $\sqrt{s} = 240$\,GeV is significantly worse than the ultimate statistical precision of 0.1\% (0.2\%) that could be naively hoped for with one million (200,000) ZH events at $\sqrt{s} = 240$\,GeV (365\,GeV). The opportunities and challenges to improve the experimental accuracy are examined in the following sections. The physics motivations to do so are numerous. First, the measurement of the ZH cross section provides an absolute determination of $g_{\rm HZZ}$, which in turn fixes all the other Higgs boson couplings (and its width), be they measured at FCC-ee (Eq.~\ref{eq:crossb}) or at a hadron collider (LHC, FCC-hh) where only coupling ratios can be inferred without theoretical assumptions.  A more accurate $\sigma_{\rm ZH}$ measurement therefore reduces the corresponding parametric uncertainty on all Higgs boson properties. Second, the quantum corrections to Higgs couplings are at the level of a few \% in the Standard Model (SM). The quantum nature of the Higgs boson can therefore only be tested if the measurement of its properties is pushed well below this level of precision, to a few per mil or better. For example, an improvement by a factor two of the ZH cross section measurement would turn automatically to a twice better Higgs self-coupling determination, and would enable the first discovery of this long-sought coupling.  
Lastly, interactions between the Higgs boson and other new particles at a higher energy scale $\Lambda$ typically modify the Higgs boson couplings to SM particles by less than 5\% for $\Lambda=1$\,TeV, with a $1/\Lambda^2$ dependence. A sub-per-mil accuracy on a given coupling measurement, e.g., $g_{\rm HZZ}$, would be needed to access the  $\Lambda=10$\,TeV energy scale. An analysis of the deviation pattern among all couplings would shed light on the underlying new physics.

The measurement of the Higgs boson mass, $m_{\rm H}$, with uncertainties much below 0.1\% has not been considered a priority so far in the current experimental and theoretical landscape: on the one hand, SM precision electroweak observables depend, via radiative corrections, only logarithmically on this quantity; and on the other hand, the current LHC experimental precision on Higgs coupling and width measurements is not particularly demanding in this respect for a sound theoretical interpretation. At FCC-ee, however, a measurement of $m_{\rm H}$ with the current $\pm125$\,MeV precision would translate into a parametric uncertainty of, e.g.,  0.7\% on $\rm \sigma_{\rm ZH} \times BR(H \to b\bar b)$, which is more than twice as large as the expected statistical precision on this observable (Table~\ref{tab:HiggsMeasurements}). It is estimated~\cite{deBlas:2019rxi} that a 0.01\% precision on $m_{\rm H}$, i.e., $\mathcal{O}(10)$\,MeV, would be enough to predict Higgs boson absolute production cross sections and decay branching fractions with an accuracy sufficiently smaller than their corresponding expected statistical precision. This requirement is almost met by the precision of 20\,MeV ultimately reachable at HL-LHC~\cite{Cepeda:2019klc}. One notable exception is the determination of the electron Yukawa coupling via resonant $\rm e^+e^- \to H$ production at $\sqrt{s} = 125$\,GeV, for which a precision smaller than the SM Higgs total width of 4.1\,MeV is required~\cite{dEnterria:2021xij}. 

The opportunities and challenges to achieve the relevant precisions on the Higgs boson mass and production cross section at FCC-ee are now examined.

\section{Opportunities and challenges: The ``recoil mass'' method}
\label{section:x-section}

The precise determination of the Higgs boson coupling to the Z boson and of the Higgs boson mass at an ${\rm e^+e^-}$ Higgs factory is  initially optimised as follows.
\begin{enumerate}
\item The centre-of-mass energy is chosen so as to maximise the number of ZH events. At FCC-ee, the luminosity steeply increases as the centre-of-mass energy decreases, so that the centre-of-mass energy was fixed to 240\,GeV, approximately 15\,GeV below the value that maximises the theoretical ZH cross section~\cite{Gomez-Ceballos:2013zzn}. 
\item In an initial approach, only the leptonic decays of the Z boson (${\rm Z \to \ell^+\ell^-}$, with $\ell = {\rm e}$ or $\mu$) are used for the cross-section measurement, as they allow the ZH events to be inclusively and efficiently selected independently of the Higgs boson decay mode. This choice is therefore effective towards an almost fully model-independent determination of the HZZ coupling, but the small Z dielectron and dimuon branching ratios are expensive in terms of statistical precision (Table~\ref{tab:HiggsMeasurements}). 
\item The mass $m_\mathrm{recoil}$ recoiling against the lepton pair is determined from total energy-momentum conservation as $m_\mathrm{recoil}^2 = s + m^2_{\ell\ell} - 2\sqrt{s}(E_{\ell^+} + E_{\ell^-})$, where $m_{\ell\ell}$ is the lepton pair invariant mass, and $E_{\ell^+}$, $E_{\ell^-}$ are the two lepton energies. In absence of initial state radiation and beam-energy spread, and with a perfect determination of the lepton pair kinematics, $m_\mathrm{recoil}$ coincides exactly with the Higgs boson mass. In practice, the Higgs boson mass and the ZH total cross section are fitted from the actual experimental $m_\mathrm{recoil}$ distribution.  
\end{enumerate}

\begin{figure}[htb]
\centering
\includegraphics[width=0.49\textwidth]{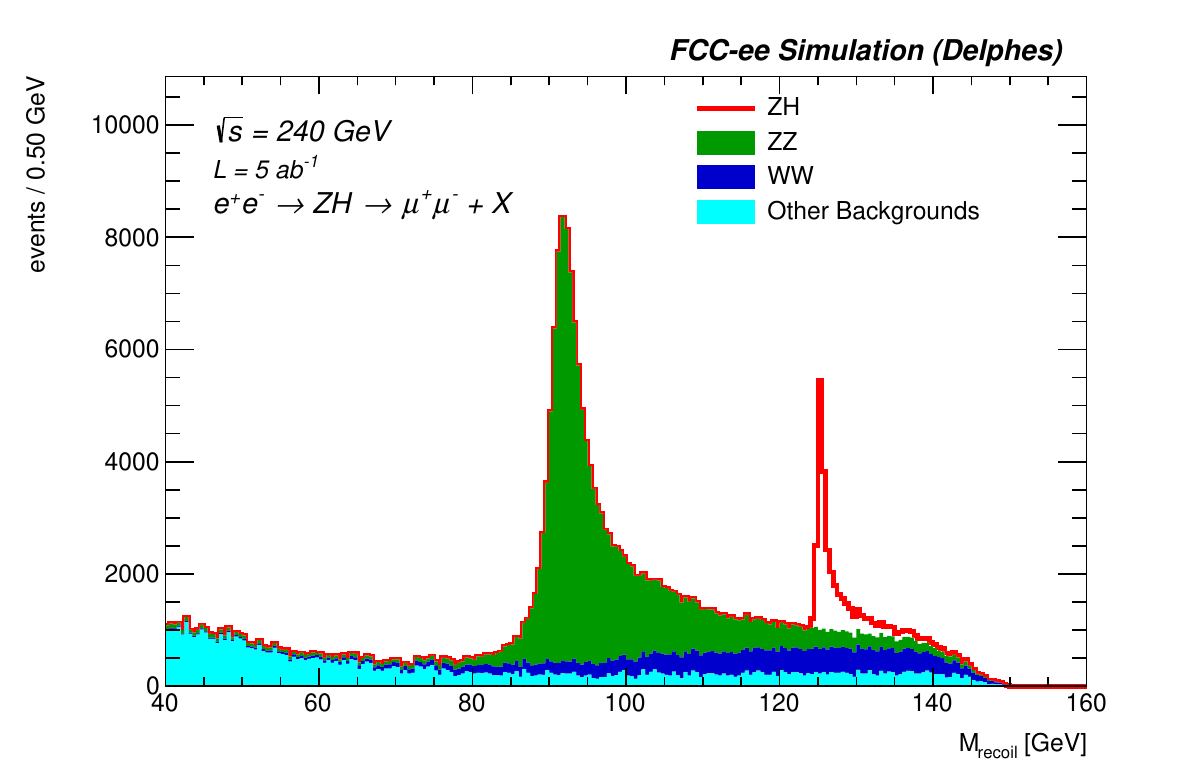}
\includegraphics[width=0.49\textwidth]{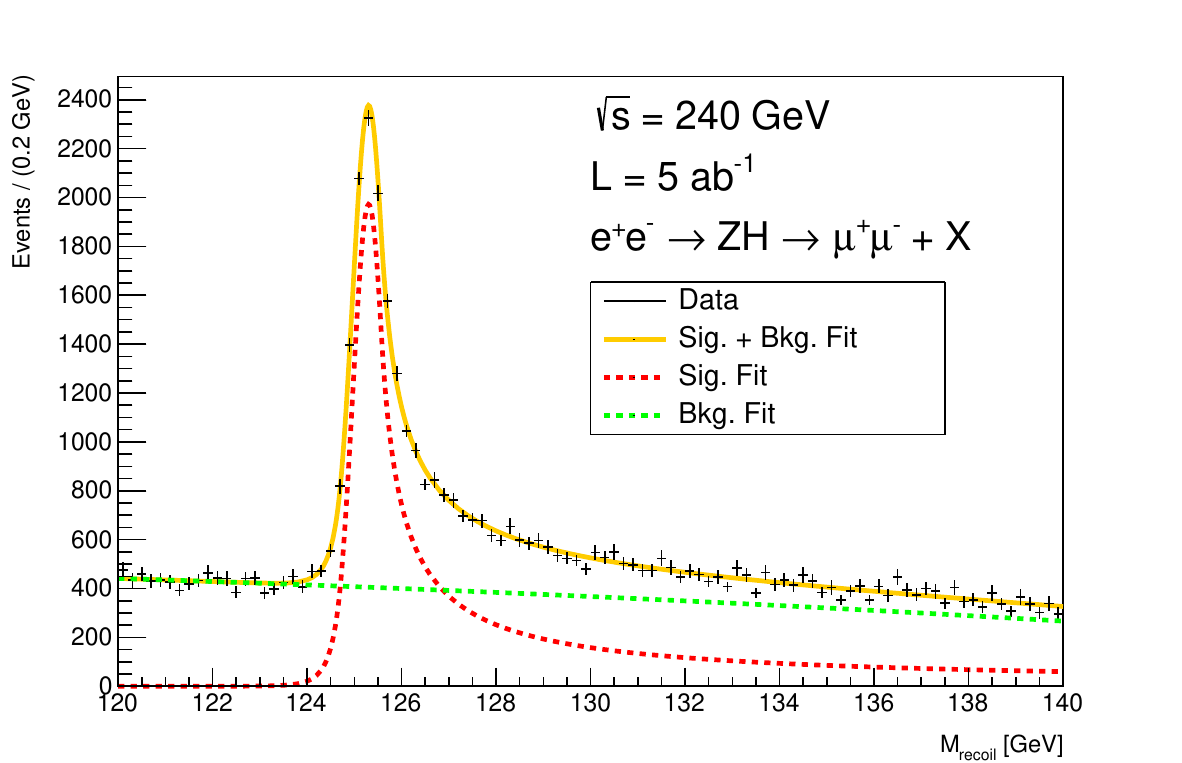}
\vspace{0.3cm}
%\end{minipage}
\caption{\small Left: Inclusive $m_{\rm recoil}$ distribution for events with a Z decaying to $\mu^+\mu^-$, between 40 and 160\,GeV displaying the Z peak from the ZZ background and the H peak from the ZH signal. Right: Expanded scale showing the $m_{\rm recoil}$ distribution in the region around $m_{\rm H}$. The ZH signal is fitted to a double-sided Crystal Ball function~\cite{DSCB,Skwarnicki:1986xj}, and the simulated background to a second-order polynomial. }
\label{fig:mrecoil}
\end{figure}

Candidate ZH events where the Z boson decays to $\mu^+\mu^-$ are selected by identifying two muons with an invariant mass close to $m_{\rm Z}$ and a total momentum transverse to the beam axis typically between 15 and 70\,GeV, while using as little information as possible from the rest of the event. The resulting $m_\mathrm{recoil}$ distribution, obtained with a DELPHES simulation~\cite{Selvaggi:2014mya} of the IDEA detector concept~\cite{Benedikt:2651299}, in particular its drift chamber~\cite{Tassielli:2021rjk}, is displayed in the left panel of Fig.~\ref{fig:mrecoil} for an integrated luminosity of $5\,{\rm ab}^{-1}$ simulated at $\sqrt{s}= 240$\,GeV and with a nominal Higgs boson mass of $m_{\rm H} = 125$\,GeV. The  background processes include the dominant diboson production $\rm e^+e^- \to WW$ and ZZ (where ``Z'' can be a Z or a virtual photon), the single boson production $\rm e^+e^- \to Z e^+e^-$, as well as the (radiative) dilepton events ${\rm e^+e^-} \to (\gamma) \ell^+\ell^-$. The dilepton and diboson background processes were simulated with {\tt Pythia}~\cite{Sjostrand:2007gs}, while {\tt WHIZARD} was used for the other background processes and the signal~\cite{Kilian:2007gr}.

%The main point in these analyses is that the events are selected only based on one reconstructed Z, with an invariant mass close the Z mass, and a $p_T$ of the Z candidate typically greater than 10 GeV or so, and not using information from the rest of the event. 

In the right panel of Fig.~\ref{fig:mrecoil}, the recoil mass distribution is fitted around $m_{\rm H}$ with a double-sided Crystal Ball function for the signal and a $2^{\rm nd}$-order polynomial for the background. To minimise the biases and the need for a-posteriori corrections arising from this choice of specific functional forms, the Higgs boson mass and the ZH cross section can also be adjusted from template distributions obtained from simulation, and calibrated with control processes with data. %In both cases, given the excellent S/B ratio, the precision obtained ($\sim 1\%$) on the number of the signal events approaches the statistical precision of the ZH sample, which is around 0.7\% for the $\mupmum$ analysis.  
%For the Z to jet-jet channel, the benefit of the higher statistics is mitigated by the less precise event reconstruction, but after combination of the three main channel ($ee$, $\mu\mu$ and jet-jet) a final precision on the inclusive cross section below 1\% is expected, and, as already stressed this result will be independent of the Higgs boson decay pattern.
%Furthermore, from the peak position in the $m_\mathrm{recoil}$ distributions in the leptonic channels, it will be possible to have a precision on the Higgs boson mass of 4 MeV. 
%The width of the Higgs cannot be obtained from the width of the resonance since it is dominated by the beam energy spread, by ISR effects, and by energy/momentum resolution of the detector if the Higgs boson width is close to 4.2 MeV as predicted in the SM. However, The measurement of the H$\to $ZZ decay rate provides the total Higgs width, and the Higgs boson decay branching ratios can be determined by measuring the ZH cross sections of the individual Higgs boson decay modes and relating them to the inclusive ZH cross section. 
%All these precise measurements will provide sensitive probes to potential new physics beyond the SM.

In both cases, and even in this first simple approach, the large signal-to-background ratio on the one hand, and the excellent drift-chamber muon momentum resolution on the other, offer the possibility to determine the inclusive ZH cross section and the Higgs boson mass with a statistical precision of $\sim$1\% and $\sim$6\,MeV, respectively. 
The muon momentum resolution achieved with the CLD Silicon tracker~\cite{Bacchetta:2019fmz} is affected by its larger amount of material, and therefore of multiple scattering, than in the IDEA drift chamber. The effect on the ZH cross section determination is marginal (given the large signal-to-background ratio), but it directly translates to a degraded statistical precision of 9\,MeV on the Higgs boson mass. This observation will need to be included in the requirements on the detector design, if a precision of $\mathcal{O}(1)$\,MeV is to be achieved on $m_{\rm H}$. 

Experimental systematic effects are also expected to be relatively larger for the mass than for the cross section, when compared to the corresponding statistical precision -- a fraction of a per cent for $\sigma_{\rm ZH}$ and a few $10^{-5}$ for $m_{\rm H}$. Methods to tackle and calibrate these effects will therefore need to be carefully designed. First and foremost, the centre-of-mass energy (and its spread) -- which enters directly in the calculation of the recoil mass -- must be determined with a similar or better accuracy. The requirements on the detector design to achieve such a precision on $\sqrt{s}$, regarding in particular the lepton and jet angular resolution, as well as systematic detector acceptance and possible hadronic effects, can be studied with a consolidated analysis of the ${\rm e^+e^- \to Z}(\gamma)$ process ($Z \to \ell^+\ell^-$, ${\rm q \bar q}$) at $\sqrt{s} = 240$\,GeV, as proposed in Ref.~\cite{Blondel:2019jmp}, with realistic FCC-ee collision parameters. The feasibility of a calibration of the method, to reduce systematic uncertainties of various origins, with ${\rm e^+e^- \to Z}(\gamma)$ events recorded at the WW threshold -- where the centre-of-mass energy can be determined with resonant depolarization with a few 100\,keV accuracy as well -- will need to be ascertained. The centre-of-mass energy spread can be inferred and monitored with an analysis of dimuon events as explained in Ref.~\cite{Blondel:2019jmp}. The absolute muon momentum scale -- and its stability -- is the second essential input to the determination of $m_{\rm H}$. The need of calibration data around the Z pole ($\sqrt{s} \simeq 91.2$\,GeV), recorded with a regular frequency, has to be estimated in this respect, complemented by the (radiative) dimuon final state and the $\rm e^+e^- \to ZZ \to \ell^+\ell^- X$ process at $\sqrt{s} = 240$\,GeV. The latter can also be exploited to check the shape of the ZZ background and tune the Monte Carlo generators accordingly. 

Several avenues should be explored to improve the $m_{\rm H}$ precision to the desired level. The possibility to increase the experiment magnetic field from 2 to 3\,T, which directly improves the momentum resolution by 30\%, will be evaluated. This study includes checking that a reasonable luminosity can be preserved in this configuration for the Z calibration data, in spite of a significant beam emittance blow-up at the interaction point. The $\rm Z \to e^+e^-$ decays might boost the precision to almost 4\,MeV with the IDEA drift chamber, but additional work is in order in this channel with a heavier tracker, for which a dedicated Bremsstrahlung photon recovery will be needed to preserve the recoil mass resolution. Experience with a full simulation of the CMS detector~\cite{Azzi:2012yn}, with a much heavier tracker than the CLD concept, showed that a performance with electrons similar to that with muons can ultimately be achieved. As also demonstrated in Ref.~\cite{Azzi:2012yn}, most of the exclusive final states, with specific Z decays ($\ell^+\ell^-$, $\tau^+\tau^-$, $\nu\bar\nu$, $\rm q\bar q$) and H decays ($\rm b\bar b$, $\tau^+\tau^-$, $\gamma\gamma$, $\mu^+\mu^-$) provide an estimate of the Higgs boson mass on an event by event basis, either from the total energy and momentum conservation constraints, or from the direct mass resolution in the case of the $\rm H \to \gamma\gamma$ and $\mu^+\mu^-$ decays. With the CMS detector, a 30\% statistical improvement on the mass precision was shown to be possible by combining these channels to the ${\rm Z}\to \ell^+\ell^-$ decays. Taken at face value, this might bring the precision down to 2.5\,MeV with the baseline run plan, and maybe 2\,MeV with a larger magnetic field. The description of the challenges, systematic studies, and detector requirements associated to these multiple final states is beyond the scope of this short essay, and will be part of the scientific outcome of the forthcoming FCC feasibility study. The comprehensive work and intellectual input required to bring these studies to completion will be instrumental for the training of young physicists (the future leaders of the field when FCC-ee is in operation mode) to ${\rm e^+e^-}$ collider physics and techniques.

To improve the precision of the ZH cross section measurement, the inclusion of the Z decays to ${\rm e^+e^-}$ is the first obvious step.
A first challenge will then be to optimize the selection towards the best precision on $\sigma_{\rm ZH}$, or alternatively, on the Higgs self-coupling, and perform the fit parameter extraction with cutting-edge analysis methods, in order to approach the previously reported and typically 30\% more precise projections~\cite{Barklow:2017suo} used in Table~\ref{tab:HiggsMeasurements}. This optimization includes the choice of the centre-of-mass energy, which affects the total number of events~\cite{Gomez-Ceballos:2013zzn}, the lepton momenta and their resolution, and the relative enhancement of the ZH cross section from the Higgs self-coupling (Fig.~\ref{fig:HHH}), all favouring a $\sqrt{s}$ value slightly below 240\,GeV. The inclusion of the hadronic Z decay in the recoil mass method is even tougher a challenge. On the bright side, the Z hadronic branching fraction is about ten times larger than its dielectron and dimuon counterpart, which increases significantly the statistical power of the method. On the other hand, the clean identification of a $\rm Z \to q\bar q$ decay independently of the Higgs boson decay -- necessary for a model-independent measurement of the cross section -- is complicated by the ambiguities in jet clustering algorithms, the multiple possibilities to associate jets to the Z or the Higgs decays (especially when the Higgs boson decays hadronically too, which happens most often), the selection efficiency dependence on the Higgs boson decay channel, and the hadronic mass intrinsic detector resolution, which is usually significantly worse than lepton momentum resolution. An excellent hadronic mass resolution, however, would significantly reduce the other ambiguities: this observation will define the performance of the whole detector in terms of particle-flow reconstruction. (The value of the experiment magnetic field is relevant here as well.) Several attempts have been made in the past to apply the recoil mass method to hadronic Z decays, the most thorough of which can be found in Ref.~\cite{Thomson:2015jda}. A fresh and systematic look is now in order, to bring in new ideas, to test cutting-edge analyses methods, and to develop innovative detector concepts, motivated by the first potential discovery of the Higgs self-coupling at FCC-ee.

%In the FCC-ee context, the requirements on the detector design (electron energy and muon momentum resolution, in particular) to achieve a few MeV statistical precision have been checked in these channels (Section~\ref{section:x-section}). The feasibility of a calibration of the method --to reduce systematic effects due to \eg\ momentum scale determination and stability-– can be evaluated with the $\epem \to \PZ\PZ \to \ell^+\ell^- X$ process. The Higgs boson mass can also be determined with the fully hadronic final state, $\epem \to \PZ\PH \to \qqbar\bbbar$~\cite{Azzi:2012yn}. The requirements on the detector design (b-tagging efficiency and purity, jet angular resolution), to achieve a precision on the Higgs boson mass of the same order as that obtained in the leptonic final state, will be studied in the context of a full 5C kinematic fit, as described \eg\ for the W mass determination at FCC-ee\cite{Beguin:2019tsx}. The feasibility of a calibration of the method -- to reduce systematic effects due to \eg\ final-state jet-jet interaction – can be ascertained with the $\epem \to \PZ\PZ \to \qqbar\bbbar$ process. %For both leptonic and hadronic final states, the need for calibration data at the Z pole will be estimated (frequency, number of events).

\section{Opportunities and challenges: scan of the ZH threshold}
\label{section:threshold}

\begin{figure}[htbp]
\begin{center}
\includegraphics[width=0.65\hsize]{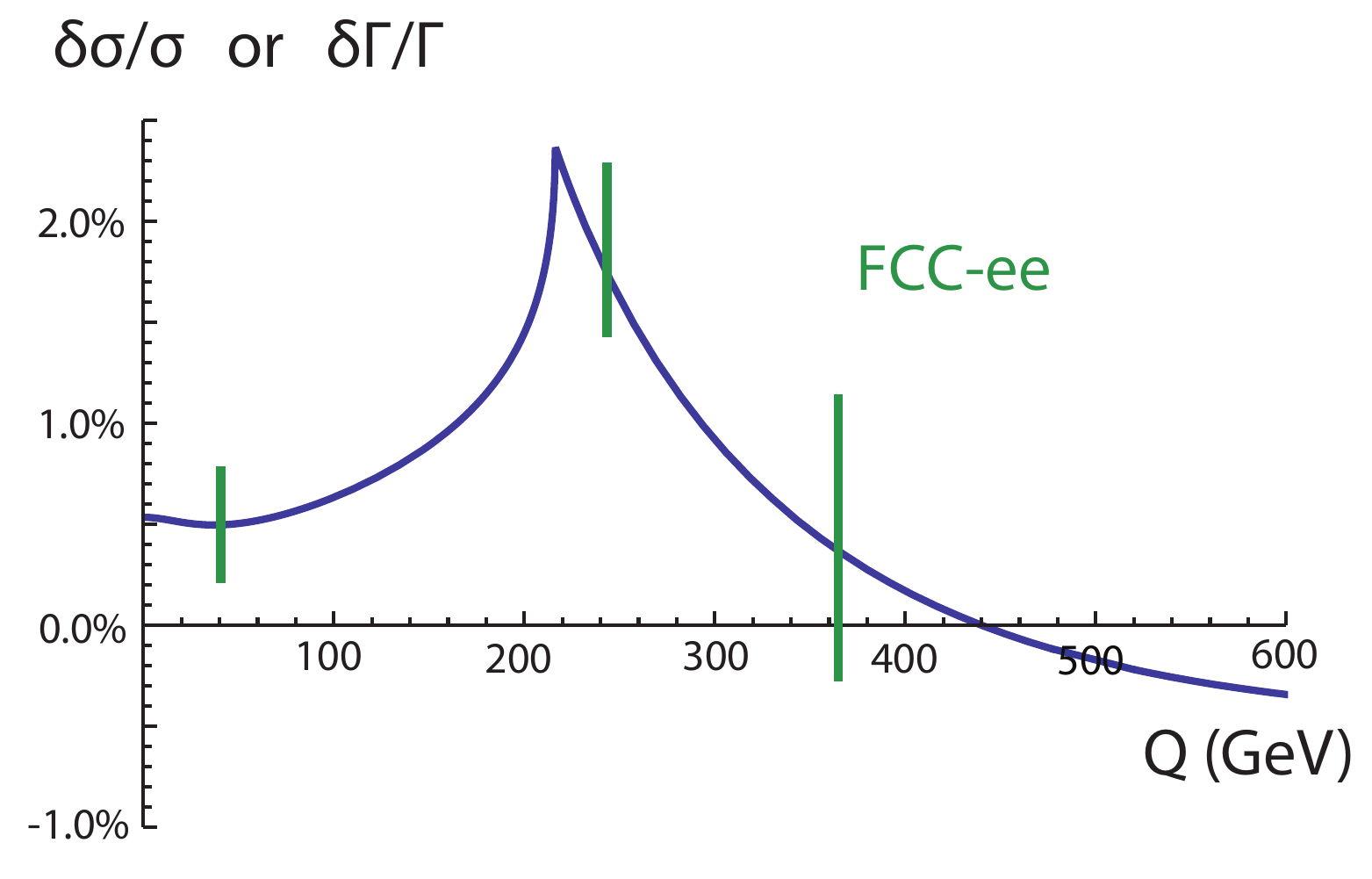}
\end{center}
\vspace*{-0.3cm}
\caption{Adapted from Fig. 9.11 of Ref.~\cite{DiMicco:2019ngk}: Relative change in the SM $\rm e^+e^-\to Z^\ast \to ZH$ Born cross section, or in the $\rm H\to WW^\ast$ partial width, as a function of $Q$ (blue curve), where $Q^2$ is the four-momentum squared of the off-shell vector boson, $\rm Z^\ast$ or $\rm W^\ast$. The change is caused by the one-loop diagrams involving the triple Higgs coupling shown in Fig.~\ref{fig:h3-TGC-atEE}. The quantity $Q$ equals $\sqrt{s} = 240$ or 365\,GeV for the ZH production cross section. For the $\rm H \to WW^\ast$  decay width, $Q$ is set to its maximum value $m_{\rm H} - m_{\rm W} \simeq 40$\,GeV. The vertical lines show the uncertainties expected from FCC-ee single measurements of these quantities. The relative enhancement is largest at the ${\rm e^+e^- \to ZH}$ production threshold, $\sqrt{s} \simeq 216$\,GeV. 
\label{fig:HHH}}
\end{figure}

In analogy to the W boson mass and width determination from the measurement of the W-pair threshold cross section lineshape~\cite{Azzi:2017iih,Beguin:2019tsx,Azzurri:2021yvl}, the cross-section lineshape for the $\Pep\Pem\rightarrow\PZ \PH$ 
process at the production threshold ($\sqrt{s} \approx m_{\rm H} + m_{\rm Z} \simeq 216$\,GeV) can be exploited to 
determine the Higgs boson mass (and width). This collision energy point is currently not foreseen in the baseline FCC-ee operation model~\cite{Benedikt:2651299}, but no stone should be left unturned. A side motivation for examining this new opportunity is illustrated in Fig.~\ref{fig:HHH}: the relative enhancement of the ZH cross section due to the Higgs self-coupling, via the loop diagrams shown in Fig.~\ref{fig:h3-TGC-atEE}, is maximal at the ZH production threshold and could be best constrained at this energy. For this threshold measurement to be useful, the Z boson mass and width need to be known with a precision significantly better than the target precision on the Higgs boson mass. The FCC-ee run at the Z pole, with its superior control of the centre-of-mass energy~\cite{Blondel:2019jmp,Blondel:2021zix}, satisfies such a requirement. The accuracy of the centre-of-mass energy at the ZH threshold (with $\rm e^+e^- \to Z(\gamma)$ events, already alluded to in Section~\ref{section:x-section}) must also be smaller than the target $m_{\rm H}$ precision. 

\begin{figure}[htbp]
\centerline{
\includegraphics[width=0.41\textwidth]{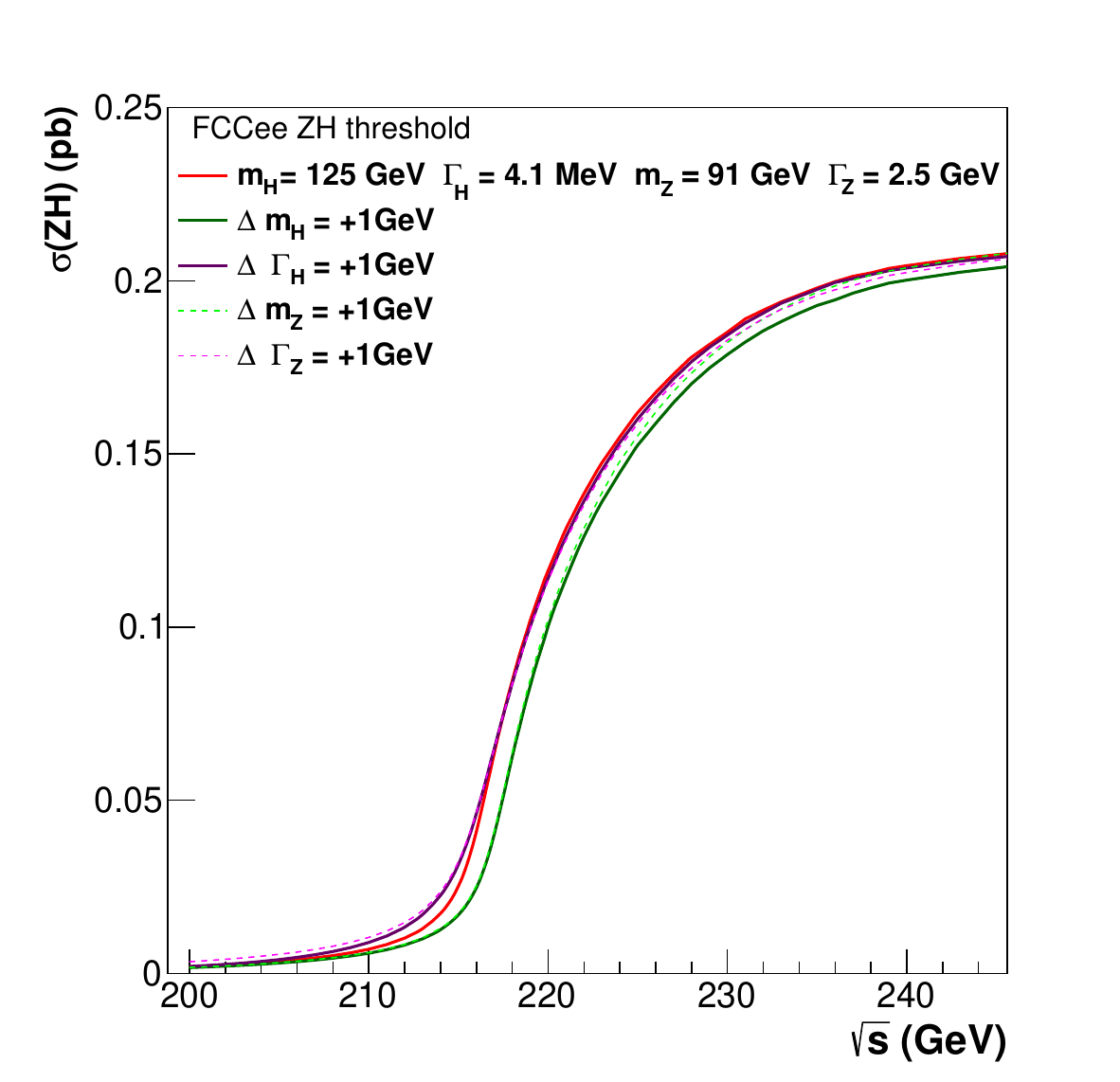}
\includegraphics[width=0.58\textwidth]{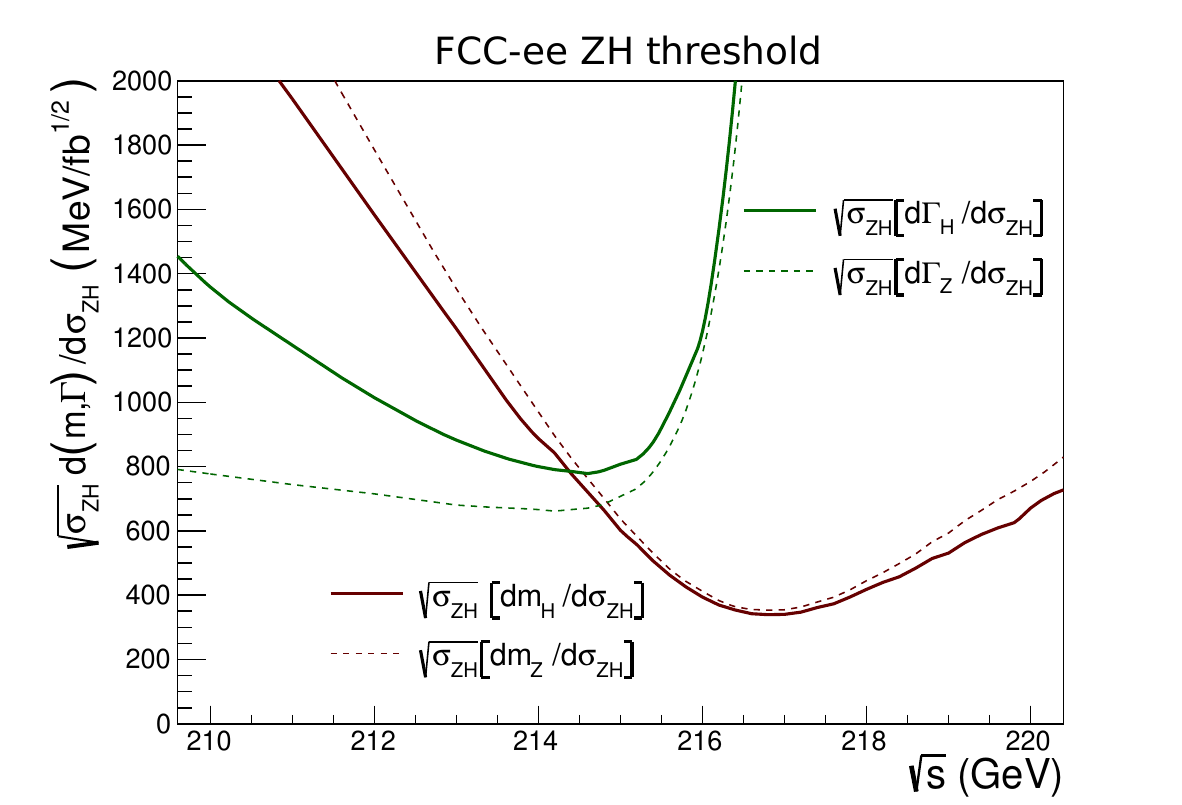}}
\caption{\label{fig:shzthr} \small Left: ZH production cross section as a function of $\sqrt{s}$ around the production threshold. The full red curve is obtained with the default Z and H masses and widths and the full green curve with a Higgs boson mass increased by 1\,GeV. The other three curves (full purple, dashed green, and dashed pink) correspond to 1\,GeV increases of the Higgs width, the Z mass, and the Z width respectively. Right: ZH cross section differential functions with respect to the Z and Higgs boson masses and widths, for an integrated luminosity of $1\,{\rm fb}^{-1}$ and a fully efficient and pure ZH event selection.}
\end{figure}

To determine the optimal centre-of-mass energy for the Higgs boson mass determination, the ZH cross section is determined here with \MGaMC~(v2.6.7)~\cite{Alwall:2014hca} as a function of the centre-of-mass energy, the Higgs boson mass and width, and the Z boson mass and width. The ZH cross section is displayed in the left panel of Fig.~\ref{fig:shzthr} as a function of $\sqrt{s}$ from 200 to 250\,GeV, with $m_{\rm Z} = 91$\,GeV, $\Gamma_{\rm Z} = 2.5$\,GeV, $m_{\rm H} = 125$\,GeV and $\Gamma_{\rm H}=$ 4.1\,MeV. For illustration, the ZH cross section lineshapes are also shown with large (1\,GeV) increases of each mass and width. For $\sqrt{s}$ around 216\,GeV, a large sensitivity to the Higgs boson mass is observed, with a variation of the cross section by a factor two. At this centre-of-mass energy, the sensitivity to the Z and Higgs boson widths is minimal. 
The statistical sensitivity to the Higgs boson mass of a $\sigma_{\rm ZH}$ measurement at a single collision energy point is given by
\begin{equation}
  \Delta m_{\PH} {\rm (stat)} =
    \left( \frac{d\sigma_{\rm ZH}}{dm_{\PH}}\right)^{-1}
    \sqrt{\frac{\sigma_{\rm ZH}}{{\cal L} \epsilon p}} = 
    \left( \frac{d\sigma_{\rm ZH}}{dm_{\rm W}}\right)^{-1}
    \sqrt{\frac{\sigma_{\rm ZH}}{\epsilon \cal L}} \sqrt{1+\frac{\sigma_B}{\epsilon \sigma_{\rm ZH}}} 
\end{equation}
where ${\cal L}$ is the integrated luminosity collected at this energy, $\epsilon$ is the selection efficiency, and $p$ the selection purity, alternatively expressed in terms of $\sigma_B$, the total selected background cross section. If several final states are combined, each with its efficiency $\varepsilon_i$ and purity $p_i$, the selection quality factor $Q=\sqrt{\epsilon p}$ is to be replaced by $Q=\sqrt{\sum \epsilon_i p_i}$. This purely statistical differential factor is shown in Fig.~\ref{fig:shzthr} (right), for an efficiency and a purity of 100\%, and an integrated luminosity of $1\,{\rm fb}^{-1}$.  
The statistical accuracy on $m_{\rm H}$ reaches a minimum at $\sqrt{s} \simeq m_{\PZ}+ m_{\PH} + 0.6\,{\rm GeV} \simeq 217$\,GeV, and at this point, it amounts to 
\begin{equation}
    \left( \sqrt{\sigma_{\rm ZH}} \frac{dm_{\PH}}{d\sigma_{\rm ZH}} \right) _{\min} \simeq 350\,{\rm MeV}\sqrt{{\rm fb}^{-1}} \simeq 10 \,{\rm MeV} \sqrt{{\rm ab}^{-1}},
\end{equation}
{\it i.e.}, just about a factor 10 larger than the 1\,MeV$\sqrt{{\rm ab}^{-1}}$ accuracy predicted for the W mass precision at the WW production threshold. The difference arises from both the overall smaller ZH cross section and its slower rise with collision energy. It is also interesting to note that, not unlike at the WW threshold~\cite{Azzurri:2021yvl},  the small sensitivity of the cross section to the H and Z widths vanishes at the point of maximal sensitivity to the masses.

%\begin{figure}[htbp]
%\centerline{
%\includegraphics[width=0.49\textwidth]{figures/dZH.pdf}
%\includegraphics[width=0.49\textwidth]{figures/dsZH.pdf}}
%\caption{\label{fig:mHthr} ZH cross section differential functions with respect to the Z and Higgs boson mass and width : absolute (center) and statistical (right). }
%\end{figure}

Taking this formula at face value, an integrated luminosity of 5\,ab$^{-1}$ at $\sqrt{s} \simeq 217$\,GeV would turn into a measurement of $m_{\rm H}$ with a statistical precision of about 5\,MeV. More realistically, and still optimistically assuming a selection quality factor of 0.3 (efficiency and purity values of 90\% with the leptonic recoil and 25\% with the hadronic recoil are typically achieved with Monte Carlo studies), the statistical precision would already degrade to 9\,MeV. The hadronic recoil analysis might prove very challenging at this centre-of-mass energy, with the Higgs and Z bosons produced at rest, and it is safer to assume a precision of 10\,MeV with the ${\rm Z}\to\ell^+\ell^-$ channels alone. Propagated systematic uncertainties from the centre-of-mass energy determination (between 1 and 2\,MeV), from the Z mass and width knowledge (negligible), from the knowledge of the residual background (supposedly well predictable from control processes), from integrated luminosity (2\,MeV if measured and predicted with a per-mil accuracy), and from theory (again 2\,MeV if the ZH cross section can be predicted with a per-mil accuracy), seem to be manageable but need to be estimated accurately. Of course, any new physics modifying, e.g.,  the Higgs boson coupling to the Z boson would lead to unpredictable effects.

A significant fraction of these statistical and systematic setbacks in the determination of $m_{\rm H}$ can be alleviated by using exclusive ZH channels, rather than using the recoil method, to measure $\sigma_{\rm ZH} \times \mathcal{B}({\rm Z \to f\overline f}) \times \mathcal{B}({\rm H \to X\overline X})$ at $\sqrt{s} = 217$\,GeV. On the one hand, most of the ZH events can be included with a much better purity in each specific channel, which moves the quality factor much closer to unity; and on the other, each measurement can be divided by the corresponding measurement at $\sqrt{s} = 240$\,GeV, to obtain a ratio  
\begin{equation}
R = \frac{\sigma_{\rm ZH} \times \mathcal{B}({\rm Z \to f\overline f}) \times \mathcal{B}({\rm H \to X\overline X}) \vert_{\sqrt{s}=217\,{\rm GeV}}}
{\sigma_{\rm ZH} \times \mathcal{B}({\rm Z \to f\overline f}) \times \mathcal{B}({\rm H \to X\overline X}) \vert_{\sqrt{s}=240\,{\rm GeV}}}
=  \frac{\sigma_{\rm ZH}( \sqrt{s}=217\,{\rm GeV}) } {\sigma_{\rm ZH}( \sqrt{s}=240\,{\rm GeV}) },
\label{eq:R}
\end{equation}
strictly independent of the Higgs boson branching fractions and, at least at tree level, independent of the Higgs boson coupling to the Z. Other systematic experimental and theoretical uncertainties may also delicately cancel in the ratio.  In addition, the sensitivity of $R$ to the Higgs boson mass is only slightly smaller than that of the ZH cross section itself, because $\sigma_{\rm ZH}( \sqrt{s}=240\,{\rm GeV})$ depends only mildly on $m_{\rm H}$.
%Figure~\ref{fig:mHR240} left shows the relevant differential factor for extracting $m_{\PH}$ from $R_{240}$. Uncertainties are minimum at very similar collision energies as shown in Fig.~\ref{fig:shzthr}, i.e. at $E_{\rm CM}=216-217$~GeV. 
\begin{figure}[tbh]
\centerline{
\includegraphics[width=0.62\textwidth]{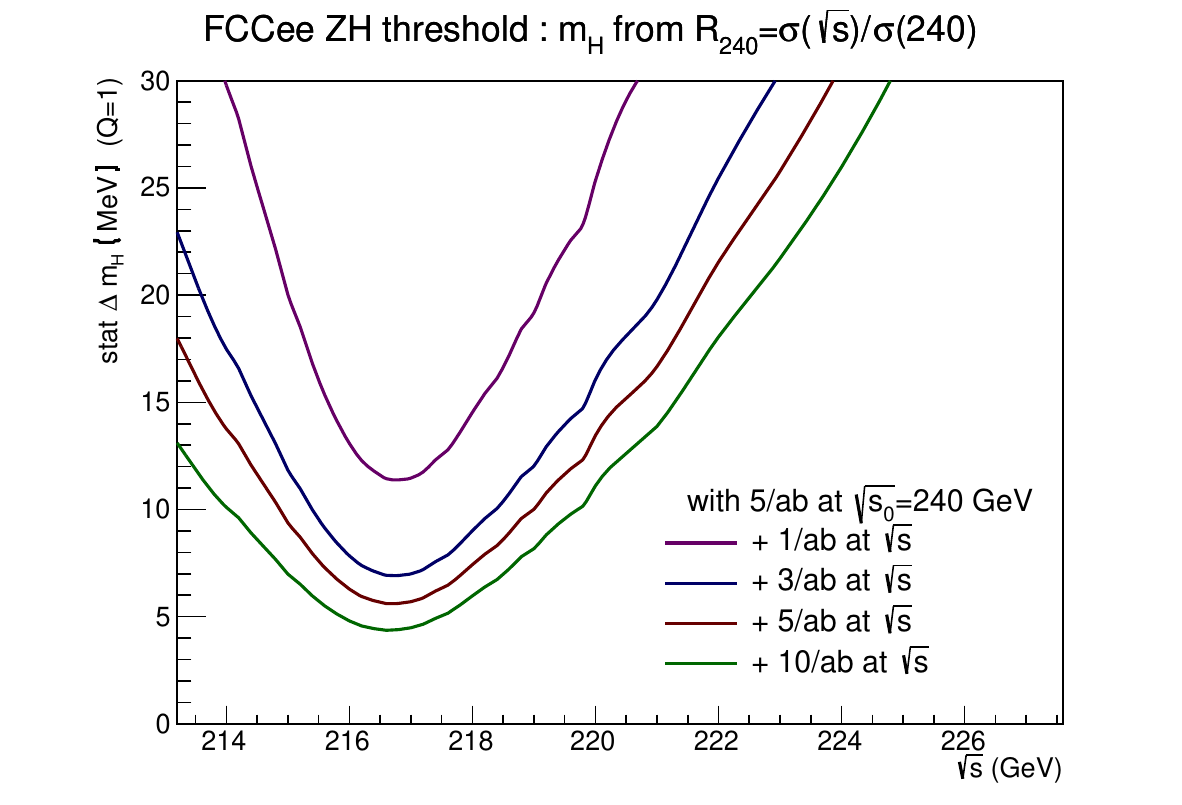}}

\caption{\label{fig:mHR240} 
%(left) Differential factors relevant for the extraction of the Higgs boson mass from the ratio $R_{240}$  of the ZH threshold cross section to the ZH cross section at $E_{\rm CM}=240$~GeV. 
\small Expected ideal statistical uncertainties on the Higgs boson mass 
 from the threshold cross section ratio $R$ (Eq.~\ref{eq:R}) assuming an integrated luminosity of 5\,ab$^{-1}$ at $\sqrt{s}=240$\,GeV and different integrated luminosities accumulated at lower centre-of-mass energies around the ZH production threshold.}
\end{figure}
Figure~\ref{fig:mHR240} shows the statistical uncertainty on $m_{\PH}$ from the measurement of the ratio $R$ with ideal event selections in all ZH final states ($Q=\sqrt{\epsilon p} =1$), assuming an integrated luminosity of 5\,ab$^{-1}$
at $\sqrt{s}=240$\,GeV and 1, 2, 5 or 10\,ab$^{-1}$ collected at lower centre-of-mass energies. A precision very close to 5\,MeV can be obtained with 5\,ab$^{-1}$ at $\sqrt{s} = 217$\,GeV. The impact from the $Q$ factor is expected to be much milder than with the recoil method, but needs to be estimated and optimized for each of the exclusive ZH final states. This precision can then be combined with the direct $m_{\rm H}$ reconstruction in each of the exclusive ZH final states at 217\,GeV, and with that obtained at 240\,GeV, which may allow the figure of 1\,MeV to be ultimately approached. Whether collecting 5\,ab$^{-1}$ at 217 and 240\,GeV is better than collecting 10\,ab$^{-1}$ at 240\,GeV only (or whatever the optimal energy turns out to be for the Higgs self-coupling determination) remains to be demonstrated. 

\section{Conclusions}
\label{section:conclusion}

The large luminosities provided by FCC-ee at centre-of-mass energies above 200\,GeV, and the possibility to simultaneously operate different detectors at various interaction points (IPs), offer multiple opportunities and techniques to measure the Higgs boson mass and production cross section with ambitious accuracies and precisions. In the baseline FCC-ee operation model, the integrated luminosities expected to be delivered and shared between two IPs, amount to 5, 0.2, and 1.5\,${\rm ab}^{-1}$ at $\sqrt{s} = 240$, 340--350, and 365\,GeV. Exploiting these integrated luminosities, and the muon momentum resolution of the IDEA drift chamber concept, a preliminary analysis of the $\mu^+\mu^-{\rm X}$ final state with the ``recoil mass'' method leads to statistical precisions of 1\% on $\sigma_{\rm ZH}$ and 6\,MeV on $m_{\rm H}$ from the data collected at $\sqrt{s} = 240$\,GeV alone.

The inclusion of all Z and H decay channels with constrained kinematic fits of the final states can improve the $m_{\rm H}$ precision to 2--3\,MeV, for which the centre-of-mass energy calibration with the $\rm e^+e^-\to Z (\gamma)$ process needs a solid systematic uncertainty appraisal to reach or exceed this level of precision (achievable with data recorded at the WW production threshold, where beam energy calibration with resonant depolarization is available). The absolute momentum (energy) scale determination for leptons (jets) will benefit from a regular detector calibration with data produced at the Z pole. Such a precision on the Higgs boson mass would already be sufficient to make good use of a dedicated run at $\sqrt{s} = m_{\rm H}$, aiming at the determination of the electron Yukawa coupling, for which there will be much bigger hurdles on the way~\cite{dEnterria:2021xij,R0monozimmer}. 

The ``recoil mass'' method, in which the Z boson is tagged via its decay products (${\rm e^+e^-}$, $\mu^+\mu^-$, or $\rm q\bar q$) without using the information from the rest of the ZH event, offers a way to determine the ZH production cross section independently of the Higgs decay branching fractions. An absolute and model-independent determination of the Higgs coupling to the Z boson $g_{\rm HZZ}$ can therefore be accomplished from the $\sigma_{\rm ZH}$ measurement at $\sqrt{s} = 240$\,GeV, which can be used in turn as a standard candle for the measurement of all other Higgs properties. The combination of the $\sigma_{\rm ZH}$ measurements at $\sqrt{s} = 240$ and 365\,GeV opens up a simultaneous and absolute determination of the Higgs self-coupling, $g_{\rm HHH}$, with a $3\sigma$ significance in the baseline run plan. Many avenues can be explored to increase this significance, either indirectly by improving the $m_{\rm H}$ precision, thereby diminishing the correlations with $\sigma_{\rm ZH}$, or by directly reducing the $\sigma_{\rm ZH}$ uncertainty. These avenues include the refined optimization of the choice of centre-of-mass energies, and the corresponding integrated luminosity sharing; an increase of the detector magnetic field to improve the lepton and jet resolutions; the use of modern analysis methods to better separate signal from backgrounds; the development of innovative detector designs towards a more accurate global event reconstruction; the increase of the integrated luminosities, either by extending the FCC-ee high-energy runs beyond the baseline run plan, or by operating detectors at four IPs instead of two (or both); and even an additional run at the ZH threshold ($\sqrt{s} = 217$\,GeV) that would deliver an additional independent $m_{\rm H}$ measurement. Matching the experimental and theoretical accuracies to the statistical accuracy needed to reach the first $5\sigma$ discovery of the Higgs self-coupling at FCC-ee will be a fascinating journey in the coming years. 

\bibliographystyle{myutphys}
\bibliography{references}

\begin{thebibliography}{10}
\bibitem{DSCB}
 {A double-sided Crystal Ball function, named after the Crystal Ball
  Collaboration, is composed of a Gaussian distribution at the core, connected
  with two power-law distributions describing the lower and upper tails. Both
  the function and its first derivative are continuous. See Appendix E of the
  next reference for an early example.},.
\end{thebibliography}

\section*{\small Data availability}

{\small \it Raw data were generated at CERN. Derived data supporting the findings of this study are available from the corresponding author upon request}

\end{document}